%% file: main.tex
\documentclass[conference]{IEEEtran}
\IEEEoverridecommandlockouts
\usepackage{cite}
\usepackage{amsmath,amssymb,amsfonts}
\usepackage{algorithmic}
\usepackage{graphicx}
\usepackage{hyperref}
\usepackage{textcomp}
\usepackage{xcolor}
\usepackage{colortbl}
\usepackage{multirow}
\usepackage{import}
\usepackage{xspace}
\usepackage{footnote}
\usepackage[tight,footnotesize]{subfigure}
\def\BibTeX{{\rm B\kern-.05em{\sc i\kern-.025em b}\kern-.08em
    T\kern-.1667em\lower.7ex\hbox{E}\kern-.125emX}}
\input{commands}
\makesavenoteenv{tabular}
\makesavenoteenv{table}
\begin{document}

\title{\tool: a web-based framework\JDR{tool?} for defining \JDR{positional?!?} modeling editors 
}
\title{Enhancing syntax expressiveness in\\ domain-specific modelling}

\author{\IEEEauthorblockN{Damiano Di Vincenzo\IEEEauthorrefmark{1},
Juri Di Rocco\IEEEauthorrefmark{2}, Davide Di Ruscio\IEEEauthorrefmark{2}  and Alfonso Pierantonio\IEEEauthorrefmark{2}
}
\IEEEauthorblockA{Universit\`a degli Studi dell'Aquila\\
67100 L'Aquila, Italy\\
\IEEEauthorrefmark{1}damiano.divincenzo@graduate.univaq.it,
\IEEEauthorrefmark{2}name.surname@univaq.it}}

\maketitle

\begin{abstract}
Domain-specific modelling helps tame the complexity of today's application domains by formalizing concepts and their relationships in modelling languages. While meta-editors are widely-used frameworks for implementing graphical editors for such modelling languages, they are best suitable for defining {novel} topological notations, i.e., syntaxes where the model layout does not contribute to the model semantics. In contrast, many engineering fields, e.g., railways systems or electrical engineering, use notations that, on the one hand, are standard and, on the other hand, are demanding more expressive power than topological syntaxes. In this paper, we discuss the problem of enhancing the expressiveness of modelling editors towards geometric/positional syntaxes. Several potential solutions are experimentally implemented on the \tool web-based platform with the aim of identifying challenges and opportunities.

\end{abstract}


\input{01-introduction.tex}

\input{comparison_table.tex}
\section{Background and motivation}\label{sec:background}
\input{02-background.tex}
\section{The JJodel modelling environment}\label{sec:overview}
\input{03-overview.tex}

\section{Examples of domain-specific environment}\label{sec:example}
\input{04-example.tex}

\section{Related work}\label{sec:related}
\input{related.tex}
\section{Discussion and Conclusions}\label{sec:discussion}
\input{05-discussion.tex}
\bibliographystyle{bibliography/IEEEtran}
\bibliography{bibliography/main}

\end{document}

%% file: commands.tex
\usepackage{xspace}
\usepackage{amssymb}
\usepackage[colorinlistoftodos]{todonotes}
\usepackage{xcolor}
\newcommand{\nb}[2]{
	\fbox{\bfseries\sffamily\scriptsize#1}
	{\sf$\blacktriangleright$\textit{#2}$\blacktriangleleft$}
}

\newcommand*{\ie}{i.e.,\@\xspace}

\newcommand*{\etal}{\textit{et al.}\@\xspace}

\newcommand\JDR[1]{\textcolor{orange}{\nb{JDR}{#1}}}

\newcommand\JJODEL[1]{\textit{jjodel}\@\xspace}
\newcommand{\tool}{\textit{jjodel}\@\xspace}

\newcommand*\circled[1]{\tikz[baseline=(char.base)]{\color{black} 
		\node[shape=circle,draw=cyan,fill=black!10!white,inner sep=.3pt] (char) {{{\texttt\textbf #1}}};}}

%% file: 01-introduction.tex
\section{Introduction}\label{sec:introduction}
In Model-Driven Engineering~\cite{MDE} (MDE), the all-important goal of delivering better modeling tools at lower cost has led to the emergence of meta-editors to produce such tools more efficiently \cite{abrahao2017user}. Frameworks like GMF~\cite{GMF}, Eugenia~\cite{kolovos2017eugenia}, and Sirius~\cite{SIRIUS} are powerful means for building graphical editors for modeling languages based on EMF\footnote{https://www.eclipse.org/modeling/emf/}. Arguably, a contributory factor limiting the application range of such frameworks is that their support is confined to the class of topological notations in which the model layout does not contribute to the model semantics.

\begin{figure}[b!]
\vspace{-0.8cm}
    \centering
    \includegraphics[width=0.2\linewidth]{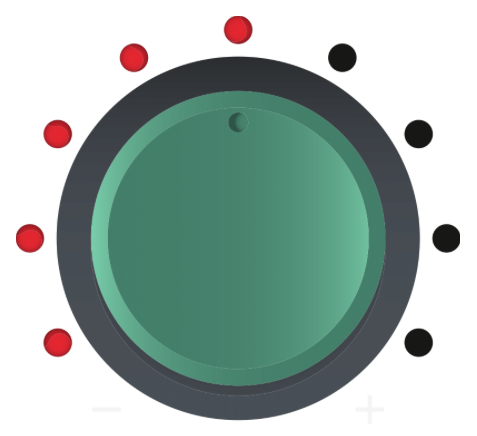}
    \caption{"Pump up the volume", an amplifier knob}
    \label{fig:knob}
\end{figure}

Conventional wisdom suggests that topological notations are sufficiently expressive for modeling the inherent information concerning software, although exceptions like sequence diagrams are still possible. Indeed, most of the time, representing knowledge using attributed nodes and relationships seems expressive enough and the most natural way of doing it. Like this, usual operations like resizing, moving, or even rotating elements in the editors are merely used for the sake of understandability and do not contribute to the definition of the model. In this respect, the correspondence between the knowledge encoded in the model and its representation can be considered \textit{unidirectional}, i.e., property dialogs are used to assign values to attributes of a model element that are then, in turn, adequately visualized using conditional formatting rules. For instance, if an attribute ranges within an interval like in the amplifier knob in Fig.~\ref{fig:knob}, current platforms do not allow to set its value by interacting with the given representation but only via a property dialog; the corresponding knob rotation is then obtained by providing a different representation for each admissible value. Such an application scenario describes a forward engineering process in which the modeling notation is designed from scratch as well as the notation syntax according to the elicited requirements provided by domain experts. In contrast, many engineering fields, e.g., railways systems or electrical engineering, use notations that are standard and cannot be modified, thus designing a modeling tool by means of the existing meta-editors may be challenging because of the lack of expressiveness of the topological notations. For instance, if the user must interact with the knob in Fig.~\ref{fig:knob} in order to assign a value to the corresponding attribute by rotating it accordingly, a \textit{bidirectional} mapping~\cite{takahashi1991general} between the layout representation and the model semantics is needed. In essence, bidirectionality permits the user to interact with the modeling elements without using property dialogs and assign attributes values that depend on layout attributes, including element coordinates, dimensions, and rotation angles. 

This paper discusses how to enhance the syntax expressiveness for domain-specific model language to go beyond topological editors. To this end, we briefly surveyed the characteristics of the existing frameworks for generating graphical editors for modeling languages. Potential solutions that have been experimentally implemented on the \tool web-based platform are analyzed and discussed through several examples to identify challenges and opportunities.

\noindent
\textbf{Outline.} The paper is structured as follows: the next section presents background and a motivational example. Section~\ref{sec:overview} illustrates the \tool modeling environment; in particular, the architecture is presented in Sect.~\ref{sec:architecture}), the syntax definition in Sect~\ref{sec:graphical_definition}, the bidirectionality between the concrete and abstract syntax is discussed in Sect.~\ref{sec:positional_feature}, the syntax layers are presented in \ref{sec:layered_syntax}, and finally how these components synergically work together is illustrated in \ref{sec:in_action}. A simple yet interesting case study in Sect.~\ref{sec:example}. Finally, Sect.~\ref{sec:discussion} draws some conclusions and prospects for future work.

%% file: comparison_table.tex

\begin{table*}[t!]
\centering
\caption{Overview of existing MDE tools providing the definition of graphical modeling workbenches}
\label{tab:comparison}
\begin{tabular}{|c|p{3.8cm}|p{2cm}|p{2cm}|p{2cm}|p{2cm}|p{2cm}|}
\hline
          &               & \textbf{\tool}                                                                      & \textbf{MetaEdit+}                                                & \textbf{GMF} & \textbf{Eugenia} & \textbf{Sirius}                                                                       \\ \hline
\multirow{4}{*}{\rotatebox{90}{\textbf{General}}}                                                       & \textbf{Environment}                                                                                                                                                                                                  & Web                                                                                                            & Standalone                                                                                                     & Eclipse                                                   & Eclipse                                                       & Eclipse                                                      \\ \cline{2-7}
\textbf{}                                                         & \textbf{Supported M3}                                                                               & ecore                                                                                                                                                                     & GOPPRR                                                                                                         & ecore                                                     & ecore                                                         & ecore                                                                                                                               \\ \cline{2-7}
                                           & \textbf{Version control system}                           & no                                                                                                                 & yes                                                                                    & any (code based)                  & any (code based)                                              & any (code based)                                                                                                                   \\ \cline{2-7}
                                          & \textbf{Model fragments}                          & no                                                                                                                                                                                  & yes                                                                                    & no                                                         & no                                                             & yes                                                                                                                                  \\ \cline{2-7} 
                                          & \textbf{Validation constraints} & yes                                                                                                            & OCL-based                   & OCL-based & OCL-based & OCL-based                                                                                                                                                                                                               \\ \hline                                            
\multirow{13}{*}{\rotatebox{90}{\textbf{Graphical syntax facilities}}}& \textbf{Multiple syntax layer}                        & yes                & yes                                                                                                     &      no                                                     &                                                  no    & yes                                                                                                \\ \cline{2-7} 
                                          & \textbf{Stack of layers}                 & yes                                                                                                                                                                                  & no                                                                                     &      no                                                     &                                       no                        & yes                                                                                                                                 \\ \cline{2-7} 
                                          & \textbf{Node customization} & complete (html/css based)                                                                                                        & predefined customization  &            predefined customization                                               &                                       predefined customization                        & predefined customization  \\ \cline{2-7} 
                                          & \textbf{Edge customization}                               & yes                                                  & yes   &      yes                                                     &                                   yes                            &                                                            yes                                                                        \\ \cline{2-7} 
                                           & \textbf{Model representation}              & any (html/css based)                                                                                             & diagram, matrix and table   &         diagram                                                  &                            diagram                                   & diagram, table, tree, sequence                                                                                                                               \\ \cline{2-7} 
                                          & \textbf{Conditional style}  & yes                                                                                                     & yes      & yes                                                       & yes                                                           & yes                       \\ \cline{2-7}

                                          & \textbf{Positional constraints}                           & yes                                                                                                                                                                                   & no                                                                                     & no & no & no                                                                                                                                                                                                                \\ \cline{2-7}
                                         & \textbf{Bidirectional semantic position}       & yes                                                                                                                                                                                   & no                                                                                     &        no & no & no                                                                                                          \\ \cline{2-7}

 & \textbf{Auto-layout}                                                                             & yes, \textit{physical} (constraint based) simulation                               & nodes alignment  &        nodes alignment                                                   &        nodes alignment                                                       & yes, with possibility to pin elements                        \\ \hline

\end{tabular}
\end{table*}

%% file: 02-background.tex
In this section, we briefly survey existing frameworks for the generation of graphical editors for modelling languages. The discussion is restricted to the only generative frameworks that, starting from a metamodel and additional attributions, can generate a fully-fledged model editor. Next, we illustrate an example of notation that the existing frameworks cannot manage without custom code extensions. 

\subsection{Frameworks for developing DSM graphical editors}
In the following, the features of a number of frameworks for generating graphical editors are analyze and discussed with the aim of outlining research challenges that can help understand where they fall short. A summary of the analysis is provided in Table~\ref{tab:comparison}.


\smallskip
\noindent\textbf{MetaEdit+}~\cite{smolander1991metaedit} is an environment for building modelling tools and generators fitting to application domains. MetaEdit+ provides a metamodelling language (\ie GOPPRR) and tool suite for defining the abstract concepts of a domain where concepts, as well as the relationships among them, are expressed by modelling infrastructure. Based on the GOPPRR, MetaEdit+ provides a set of functionalities for defining diagramming editors and model generators.
\smallskip

\noindent\textbf{GMF}~\cite{GMF} is a framework for developing graphical editors based on the Eclipse Modeling Framework (EMF) \cite{EMF} and Graphical Editing Framework (GEF).\footnote{\url{https://www.eclipse.org/gef/}}
The GMF Tooling supports a model-driven process for generating a fully functional graphical
editor based on the GMF Runtime starting from the four different models: \textit{i)} the 
\textit{Ecore-based metamodel}, \emph{ii)} the \textit{graphical definition model} that identifies graphical elements, \emph{iii)} the \textit{tooling definition} model contributes to palettes, menus, toolbars, and other periphery of the editor, and \textit{iv)} the \textit{mapping model} that establishes all links to define the links between the aforementioned models.
\smallskip

\noindent\textbf{Eugenia}~\cite{kolovos2017eugenia} aims to simplifies the development of GMF-based graphical model editors. Eugenia uses metamodel annotations to generate the four models previously mentioned by applying a well-defined model transformation chain. 
\smallskip

\noindent\textbf{Sirius}~\cite{SIRIUS} is an Eclipse project based on GMF and the EMF that is increasingly replacing GMF in building modern graphical editors because it requires little programming overhead.
Sirius allows the conceptual separation between the metamodel and the corresponding representation: starting from a Ecore-based domain specification, it allows a model-based specification of visual concrete syntax organized in viewpoints. 

\medskip
Recently, the framework Sprotty\footnote{\url{https://projects.eclipse.org/projects/ecd.sprotty}} gained some attention as it provides web-based diagramming functionalities for rendering and animating model diagrams. The framework relies on the Graphical Language Server Platform\footnote{\url{https://www.eclipse.org/glsp/}} (GLSP) and provides robust and scalable rendering experience possibilities. Unfortunately, Sprotty does not (yet?) provide meta-facilities for the generation of graphical editors, and therefore it has been omitted from the comparison with the other meta-editors.


In Table~\ref{tab:comparison}, we provide an overview of the meta-editors mentioned above in the context of graphical representation and positional semantics we discussed before. The table also includes our \tool prototype presented in Sect.~\ref{sec:overview}. The first general rows report aspects of the meta-editors, i.e., the environment where the meta-editor is run, the meta notation to define metamodels, version control systems, the support of models fragments, and the availability of model validation constraints. Then, the table reports if and how the considered meta-editors support the graphical syntax facilities.
In particular, we analyze the support of multiple syntax layers and the possibility of stacking them, the chance to define nodes and edges customization, the supported model representation, the availability of conditional style, and the auto-layout model features.
Finally, it is worth noting that apart from our \tool prototype, all the meta-editors have little or no support for positional features. In particular, positional constraint and bidirectional semantic position are only supported by \tool.




\newcommand{\code}[1]{\texttt{\small #1}\xspace}

  \begin{figure*}[t!]
    	\centering
    	\begin{tabular}{c c }	
    		\subfigure[Aircraft metamodel]{\label{fig:aircraft_class_diagram}
    			\includegraphics[width=0.4\linewidth]{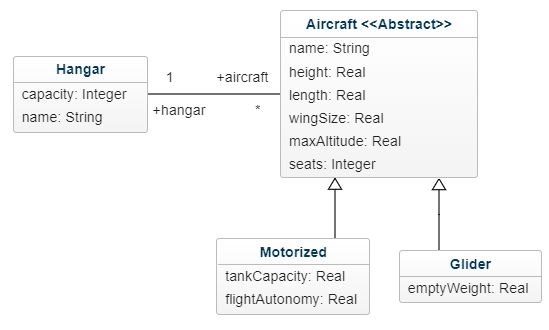}} &
    		\subfigure[An instance of the Aircraft metamodel]{\label{fig:aircraft_object_diagram}
    			\includegraphics[width=0.4\linewidth]{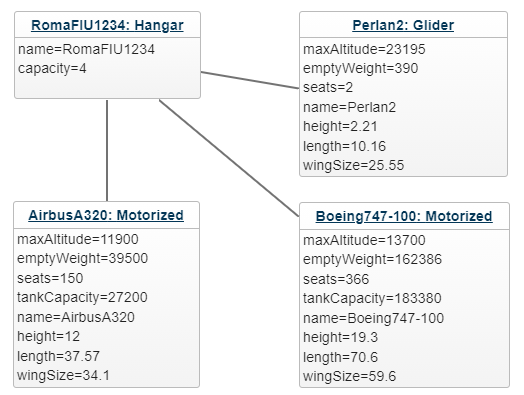}}	
    	\end{tabular} 
    	\caption{A Topological editor} 
    	\label{fig:no_topological_editor}
    \end{figure*}

    \begin{figure*}
        \centering
        \includegraphics[width=1\columnwidth]{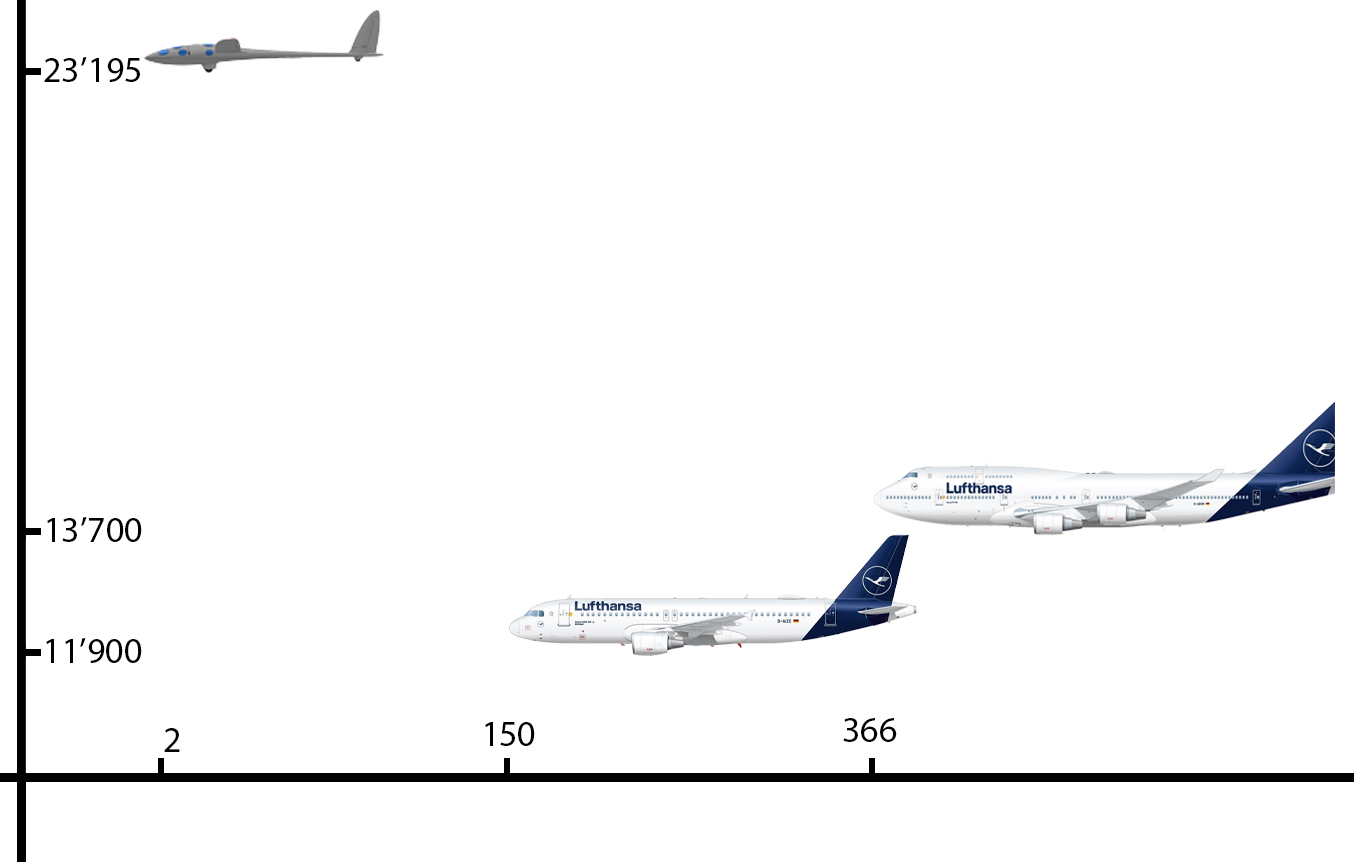}
        \caption{A positional/geometric editor.
        The diagram is approximately on scale with a Y-axis on power (0.5) scale, X-axis with linear scale, aircraft length and height on a Log2 scale, nodes are positioned by the bottom-left vertex. Images from Lufthansa}
        \label{fig:topological_editor}
    \end{figure*}  
\subsection{Motivational example}\label{sec:motivation}
This section proposes an illustrative example to show how certain representations are challenging to be managed in a topological editor. In particular, the example features a notation to represent a catalog of airplanes to be stored in a hangar. Figure~\ref{fig:aircraft_class_diagram} presents an excerpt of the metamodel formalizing such a domain, where the \code{Hangar} stores several \code{Airplanes} that are distinguished in two categories, \code{Motorized} and \code{Glider}. Each airplane is characterized by its \code{maxAltitude}, \code{height}, \code{length}, and the number of \code{seats}. Despite its simplicity, the metamodel permits identifying some of the challenges that a positional/geometric editor should address. Let us assume that an hangar named \code{ROMAFIU1234} contains two motorized airplanes and a glider, as specified in the instance model (whose abstract syntax is) given in Fig.~\ref{fig:aircraft_object_diagram}.

In order to convey more intuition to the domain expert, an alternative representation for the model is given in Fig.~\ref{fig:topological_editor} where the dimension and the position of an airplane in the diagram denotes its size, capacity, and maximal altitude, respectively. In particular, the y-axis and x-axis represent the altitude and the number of seats, respectively, whereas the dimension of the airplane gives its length and height.     

Besides representing such information on the diagram, it would be interesting to let the domain expert instantiate an airplane and specify its characteristics by dragging and resizing the graphical elements according to the \textit{semantics} given to its geometric features. Such kind of editor behavior requires a bidirectional mapping between the model content and the model representation. More precisely, interacting with the objects in the editor to define their characteristics requires a mapping from the representation to the model abstract syntax.

Missing properties can be represented in a traditional approach through node content and styling, by the layout properties of a sub-node like the \code{tankCapacity}, or not represented at all and left to other views.

Since not all the elements in the metamodel require a \textit{reverse} mapping from the positional attributes to the model abstract syntax, there is a problem about what to do with those objects not having such attributes; solutions could be either:
    \begin{itemize}
        \item[--] exclude them from the view,
        \item[--] display them with any position and size ignoring the positional semantics, or
        \item[--] display them with custom semantic following other positional rules.
    \end{itemize}
Finally, since element positioning contributes to the model definition and semantics, such elements cannot always be freely positioned in the editors. Thus, it may be of crucial relevance defining positional constraints as, for instance, in the following cases:
\begin{itemize}
        \item[--] an aircraft cannot fly below the ground level: this means that the aircraft cannot be positioned with negative value of y-axis);
        \item[--] an aircraft cannot have less than 1 passenger: in other words the aircraft cannot be positioned before the 0 value of the x-axis;
        \item[--]  the airplane height cannot exceed its length: that would be more likely to be a mistake than a design choice.
        \end{itemize}
None of the mentioned meta-editors implements a complete bidirectional mapping between the model representation and the model content or abstract syntax nor permits the specification of layout or positional constraints. Consequently, certain common features like for instance, the \textit{grid} or the \textit{snap-to-grid} function, are either provided as "canned" abstractions of behaviors for a fixed set of graphical objects or require to be manually implemented (most of the frameworks offer extension mechanisms). While the mapping from the model content is typically part of the machinery in the generated editors as the property dialog is the only way to modify specific attributes, specifying a mapping from the model representation to its abstract syntax is not supported. 

\smallskip
To a certain extent, the presented example is artificial and does not exhaustively cover all possible cases that topological editors cannot handle. However, it shows how enhancing the expressiveness of the current meta-editors would widen the general model-driven application range, especially regarding those engineering fields where tooling has a significant role.

%% file: 03-overview.tex
\begin{figure*}
    \centering
    \includegraphics[width=\linewidth]{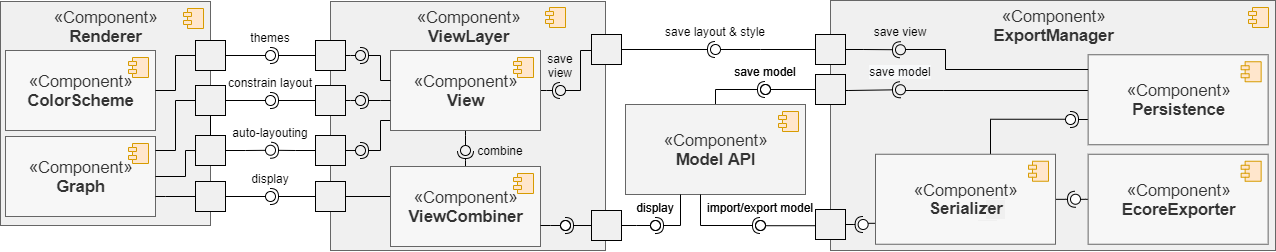}
    \caption{An overview of \tool components}
    \label{fig:component}
\end{figure*}

To mitigate the challenges proposed in Sect.~\ref{sec:background}, we propose \tool, a web-based tool for defining modelling editors. 
The architecture of \tool is presented in Sect.~\ref{sec:architecture}. 
Section~\ref{sec:graphical_definition} describes the \textit{View} components to define the graphical notation of the modelling elements.
To support the specification of positional editors, event-driven rules have been discussed in Sect.~\ref{sec:positional_feature}. Section~\ref{sec:layered_syntax} describes the possibility of stacking view layers for rendering the abstract concepts defined in the metamodel, while Sect.~\ref{sec:in_action} presents all the \tool workbench parts.

\subsection{Architecture}\label{sec:architecture}
\tool is a prototypical web-based platform that consists of four main components as illustrated in Fig.~\ref{fig:component}; in particular:
\begin{itemize}
    \item[--] the \textit{Renderer} is the component responsible for workbench arrangement as, for instance, themes and color schemes that contains two subcomponents, \textit{ColorScheme} that provides coloring customization used by \textit{Graph} that, in turn, renders the model;
    \item[--] the \textit{Model API} component provides API management to access and bridge modelling elements throughout the metamodelling architecture tiers, \ie model, metamodel, and meta-metamodel; 
    \item[--] the \textit{ViewLayer} combines different \textit{View}s and keeps the abstract syntax and the graphical concrete syntax in a consistent state; in addition, it enforces the positional semantics, with constraints and execution rules;
    \item[--] finally, the \textit{ExportManager} manages the serialization of the authored models in the XMI format\footnote{\url{https://www.omg.org/spec/XMI/2.5.1/About-XMI/}} or in JSON\footnote{\url{https://emfjson.github.io/}} by means of the \textit{Serializer} and \textit{EMFExporter} sub-components; the \textit{Persistence} sub-component is responsible for managing external repositories, e.g., MDEForge~\cite{basciani2014mdeforge} or ReMoDD~\cite{france2006repository}, where modeling artifacts can be persistently stored. 
\end{itemize}

\tool is designed to be primarily client-centered and, as such, can be used offline as the server does nothing more than serving static resources. The dynamic content, user customizations, and persistence are granted by JavaScript and the client internal \textit{localStorage}. 

\smallskip
The main objective of the architectural design is reducing the accidental complexity of the modeling tool by providing the platform with \textit{zero setup} possibilities, low or no maintenance, and simplifying the overall workflow by removing those platform-related steps like the regeneration of the editors. This has been possible by adopting the technological stack summarized in Table~\ref{tab:technologies}.

\begin{table}[h!]
\centering
\caption{Technological stack}
\label{tab:technologies}
\begin{tabular}{|p{2cm}|p{6cm}|}\hline
\textbf{Technology} & \textit{Description}\\\hline
Typescript\footnote{\url{https://www.typescriptlang.org/}} & The main programming language used for the \tool implementation\\\hline
Angular CLI\footnote{\url{https://angular.io/cli}} & The command-line interface tool that has been used to initialize, develop, and maintain the graphical elements of \tool\\\hline
jQuery\footnote{\url{https://jquery.com/}} & The JavaScript library to easily traverse and manipulate DOM documents\footnote{\url{https://www.w3schools.com/js/js_htmldom.asp}}; it has been used to manage the user-defined dynamic content\\\hline
Bootstrap\footnote{\url{https://getbootstrap.com/}} & Used to manage the UI and to aid the user in defining customized concrete syntax by well-known bootstrap shortcut classes\\\hline
vis.js\footnote{\url{https://github.com/mdaines/viz.js}} & The open source graph visualization software for representing structural information as diagrams, graphs, and networks; it has been used to handle automatic layout of the graphical elements\\\hline
\end{tabular}
\end{table}

\subsection{Graphic concrete syntax}\label{sec:graphical_definition}
\begin{figure}
    \centering
    \fbox{\includegraphics[width=0.3\columnwidth]{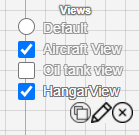}}
    \caption{The layered syntax}
    \label{fig:viewcombiner}
\end{figure}

A \tool project comprehends a metamodel and an associated specification of the graphical concrete syntax. The latter is managed by the \textit{ViewLayer} and consists of a number of layered \textit{View}s, each composed by \textit{ViewRule}s. Each \textit{ViewRule} defines the visual representation, layout constraints, execution rules, and positional semantics for one or more modelling elements through a customizable template defined by the modeler. A \textit{View} can represent the model partially or entirely and binds a modeling element at most to a single \textit{ViewRule}, although views can be combined by means of the \textit{ViewCombiner} giving place to \textit{ViewRule}s (collected from different Views) assigned to a single modelling element. Rules assigned to the same modeling element are arranged in a priority queue managed by the \textit{ViewCombiner}, which will apply the rule with the highest priority. Ultimately the \textit{ViewCombiner} can stack partial \textit{View}s solving overlaps with a priority policy as depicted in Fig.~\ref{fig:viewcombiner}.

A \textit{View} consists of a collection of \textit{ViewRules}. Each of such rules specifies the concrete syntax for one or more metaclasses by providing the HTML/SVG/CSS used in the \textit{renderer}'s \textit{graph}. Besides specifying the pictorial representation for the instances of a metamodel, rules specify also the actions to be executed when their parent view is active. When a view is activated (or deactivated) the corresponding rules are dynamically enabled (or disabled) by means of the \textit{Graph} component that displays the content computed by the \textit{View} subsystem. Like this, depending on the active views the model is given the desired pictorial representation and the editor the specified behavior as specified by the rules. It is worth mentioning that rules are not applied unconditionally but only when certain events in the editor are triggered and related conditions are verified as described in the next section.



\subsection{Bidirectional syntax mapping}\label{sec:positional_feature}
Constraints and positional semantics are enforced through the rules specified in the views. A rule consists of a triple
\[(<trigger>,<condition>,<action>)\]
where a \textit{trigger} is an event occurring on the editor that can be spontaneously activated or the outcome of the user interacting with the editor. When a \textit{trigger} is fired and the \textit{condition} is satisfied, then the \textit{action} is executed providing the editor the desired behavior. Multiple triggers can be combined in a disjunction, in case the rule can be activated by multiple events. The available triggers are listed in Table~\ref{tab:trigger} and refer to events that occur when the model content is modified, when an element is moved around, resized, or rotated. Additional triggers are fired when the mentioned actions start, end, or \textit{while} the action is executed. \textit{While} triggers are fired many times during the execution of an action.

\begin{table}[b!]
\centering
\caption{Rule triggers}
\label{tab:trigger}
\begin{tabular}{|p{2cm}|p{6cm}|}\hline
\textbf{Trigger} & \textbf{Description}\\\hline\hline
{onRefresh} & when the model data changes\\\hline\hline
{onDragStart} & when the \textit{move} action starts\\\hline 
{whileDragging} & when the \textit{move} action is being executed\\\hline 
{onDragEnd} & when the \textit{move} action ends\\\hline\hline 
{onResizeStart} & when the \textit{resize} action starts\\\hline
{whileResizing} & when the \textit{resize} action is being executed\\\hline 
{onResizeEnd} & when the \textit{resize} action ends\\\hline\hline 
{onRotationStart} & when the \textit{rotation} starts\\\hline
{whileRotating} & when the \textit{rotation} action is being executed\\\hline
{onRotationEnd} & when the \textit{rotation} ends\\\hline
\end{tabular}
\end{table}

Every time a specific trigger fires, such as changes in the model or movements on the graphical layout, if a rule with that specific trigger exists and the corresponding \textit{condition} is verified, then the \textit{action} presented in the rules will be executed. The condition can predicate on layout and model values or the outcome of other action executions. Once the condition is verified, the provided action can perform changes in the model content or its graphical representation. In addition, an action can enforce a provided constraint to make sure the positioning of the affected element is correct. The constraint enforcement is performed employing an underlying constraint solver. It is worth noting that actions always produce an output value that can be used by other actions or as a condition for triggers and optionally side effects in the layout or model content.

The actions are grouped in three main categories (for the sake of brevity, we are listing the most significant ones):

\smallskip\noindent
\textbf{Export actions}: such actions allow to propagate the computed values to other elements by editing the target element HTML content and attributes. They can be used to style a reference target according to his referrer.

\begin{figure}
    \centering
    \fbox{\includegraphics[width=0.7\linewidth]{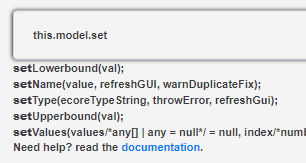}}
    \caption{context-sensitive auto completion for layout actions}
    \label{fig:autocomplete}
\end{figure}

\smallskip\noindent
\textbf{Constraint actions}: these actions prevent the user from generating an invalid layout enforcing positional semantics. Constraint actions have the following shape \textit{"property operator value"} where: \textit{i)} the \textit{property} is layout graphical attribute, e.g., the width of the current element, its absolute position or its position relative to a target; \textit{ii)}~\textit{operator} is the comparison operator; \textit{iii)}~\textit{value} is the output computed by a mathematical or JavaScript function from the modelling values, the layout properties of the current modelling element, the relative target of this rule, or the entire model. A valid example of this rule is \texttt{\small width = this.target.width * 2}, which will enforce the width of the modelling element holding this rule always to be twice the size of his target. NB: the "target" of this rule is always a single HTML target selected by a CSS selector, but it is possible to refer to multiple modelling targets accessing the model content through \texttt{\small this.model}.
Figure~\ref{fig:autocomplete} shows the completion feature to assist the user to navigate model elements;

\smallskip\noindent
\textbf{Generic action/variable actions}: they enable to access layout and modelling data and apply any valid change by using the provided API in the auto-completion editor.
They are typically used to compute an output value that other triggers or actions will use and can produce side effects and act as a fallback for actions that cannot be performed using other commands, such as constraints that cannot be expressed as a single equation.

\subsection{Layered Syntax}\label{sec:layered_syntax}
As already mentioned, a view is a collection of bundles that can customize the visual styling, layout constraints, and data constraints; such bundles will be just called \textit{style} in this sequel whenever it does not give place to ambiguity.

A single modelling element can have an arbitrary number (possibly none) of applicable styles forming a priority queue, but only the top of the queue is applied to the element. Similarly to what happens with CSS, a style can affect modelling elements according to the following priority order:

\begin{enumerate}
\item \textit{Personal}, it is applied directly to the element, and an element can only have a single personal style;
\item \textit{Inherited / inheritable}, it is inherited by the M1 from the M2, or by the M2 from M3, where M1, M2, and M3 are the usual levels in the metamodeling architecture.
If an inheritable style is defined inside an M2 classifier, all the recognized instances of the class will inherit this style.
\item \textit{Default (of a view)}, it is applied with low priority to all modelling elements.
\item \textit{Global default}, it is only applied if the queue is empty; this style is immutable and does not belong to any view.
\end{enumerate}
A view does not need to be total; it can also specify styles for only a subset of the modelling elements. Partiality allows to address specific styling needs and combine different styles as long as they do not interfere. Indeed, multiple views can be applied to the same model; however, if the same attributions are specified in multiple views, then the definitions in the view that is at the highest priority will prevail by overriding the definitions in the lower priority views. 

\subsection{\tool in action}\label{sec:in_action}

\begin{figure*}
    \centering
    \includegraphics[width=\linewidth]{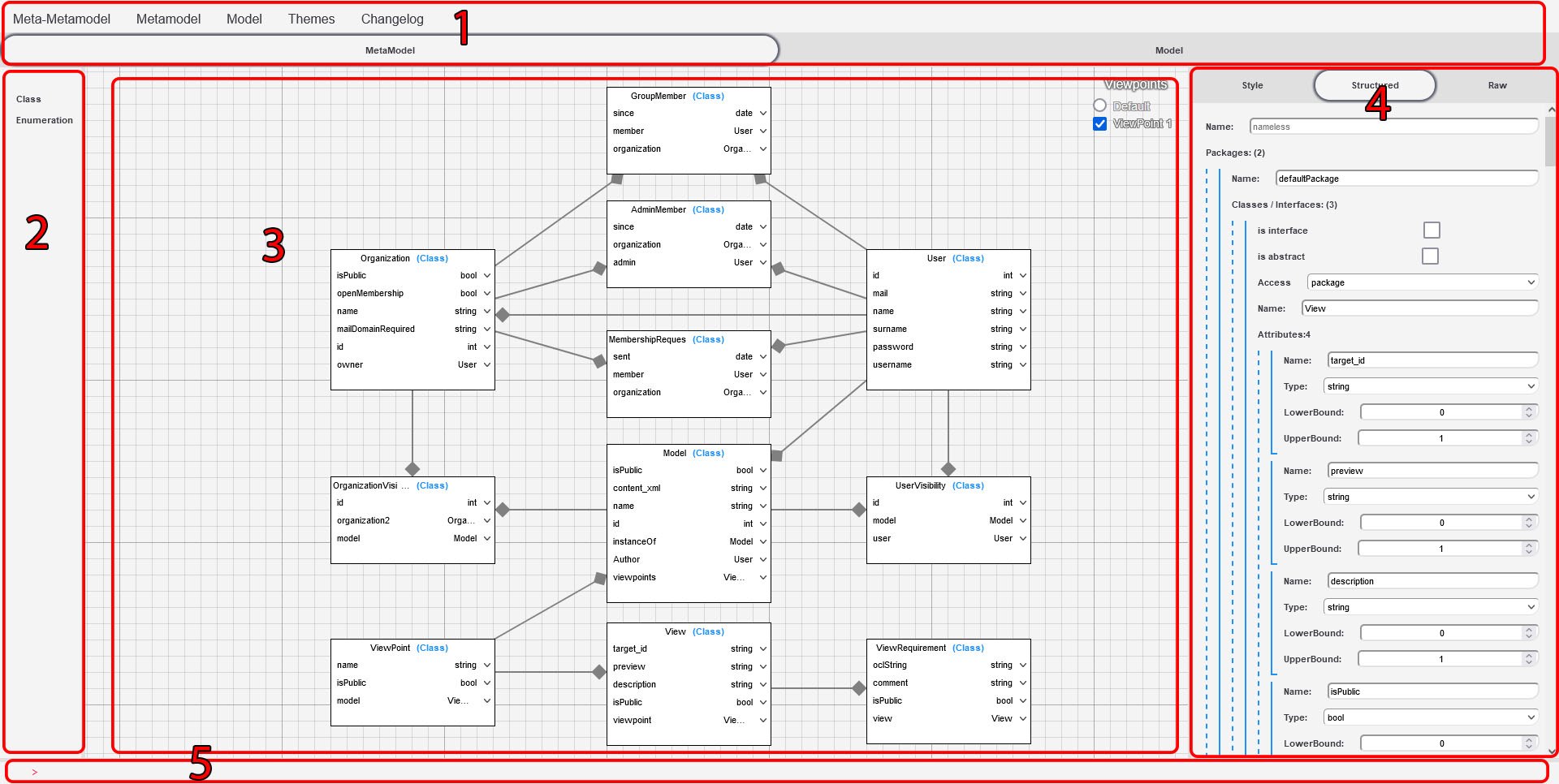}
    \caption{\tool graphical user interface}
    \label{fig:gui}
\end{figure*}
Figure~\ref{fig:gui} illustrate the \tool interface; more in detail, it presents the metamodel editor, which is based on a diagrammatic notation similar to that of the UML class diagrams. The interface consists of several panes focusing on different aspects of language definition: \textit{i)} the definition of the abstract syntax, \textit{ii)} the specification of the graphical concrete syntax for the modelling elements and \textit{iii)} the definition of both positional semantics and constraints.~More in detail
\begin{itemize}
    \item[--] the top bar \circled{1} contains app-wide or model-wide options and allows to change between metamodelling and modelling views;
    
    \item[--] the left pane \circled{2} lists the metaelement of the metamodeling notation that can be instantiated for creating or enumeration types; 
    
    \item[--] the canvas graph \circled{3} containing the model being edited (in this case, a metamodel) in the selected graphical notation; 
    
    \item[--] the right pane \circled{4} whose content is dynamic and always relative to the selected element on the graph; it composed of three subsections: \textit{i)} the \textit{structured view} is a static tree-like view of the selected model element and his sub features which can be used to edit elements not displayed in the graph, \textit{ii)} the \textit{style editor} allows to customize the graphical representation of the current modelling sub-object, and \textit{iii)} the \textit{Raw view} shows the selected element data in Ecore/JSON format;
    
    \item[--] finally, the console \circled{5} accepts queries to traversing and visiting modelling elements and programmatically updates their values (not fully implemented).
\end{itemize}

%% file: 04-example.tex
In this section, we illustrate the \tool potential by showing an implementation of the metamodel presented in Sect.~\ref{sec:motivation} and the corresponding modeling environment. In particular, the modeling environment consisting of an editor with a default view and a positional syntax is obtained from the metamodel specification and the view definition, as explained in the previous sections. The entire environment with the relative specifications is freely accessible\footnote{\url{https://bit.ly/3iK01GC}}.

\subsection{Metamodelling}
As described in the Sect.~\ref{sec:overview} the starting point of any \tool projects is the metamodel definition. The platform supports metamodeling by providing a predefined editor for specifying Ecore-based metamodels. 
\begin{figure}
    \centering
    \fbox{\includegraphics[width=0.95\linewidth]{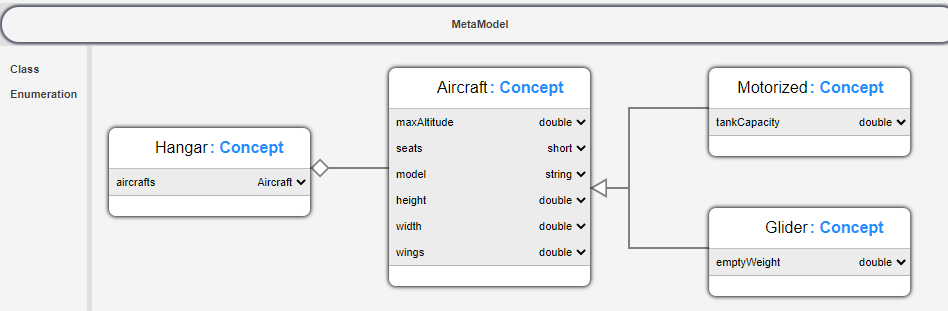}}
    \caption{The M2 default editor}
    \label{fig:4_m2}
\end{figure}
Figure~\ref{fig:4_m2} presents the \tool metamodel editor containing the example proposed in Sect.~\ref{sec:motivation}.

Once the metamodel is specified, \tool dynamically configures the corresponding model editor as described in the next section.

\subsection{Model editor}
\begin{figure}
    \centering
   \fbox{\includegraphics[width=0.25\linewidth]{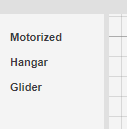}}
   \vspace{-0.2cm}
    \caption{The M1 left pane}
    \label{fig:4_m1_toolbox}
\end{figure}

The obtained model permits to instantiate the given metamodel. Thus, all the instantiable concepts (i.e., the non abstract metaclasses) formalized in the metamodel are listed in the left pane (\circled{2} in Fig.~\ref{fig:gui}). The left pane and the default view are the baselines to fully express the model content. Figure~\ref{fig:4_m1_toolbox} shows the available concepts ready to be used for authoring the model.

\subsection{Concrete syntax}
\begin{figure}
    \centering
    \fbox{\includegraphics[width=0.95\linewidth]{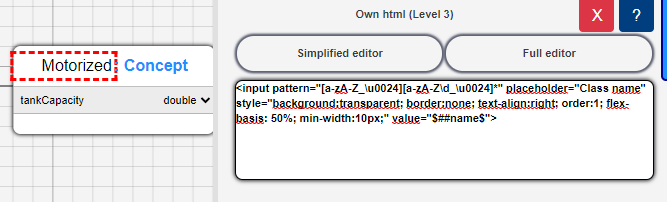}}
    \caption{A template}
    \vspace{-0.5cm}
    \label{fig:4_template}
\end{figure}

\begin{figure}[h]
    \centering
    \fbox{\includegraphics[width=0.8    \linewidth]{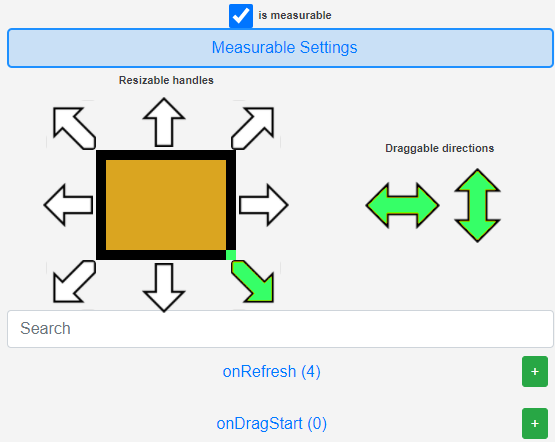}}
    \caption{Measurable settings}
    \vspace{-0.2cm}
    \label{fig:4_measurable}
\end{figure}

The default syntax provided by \tool for editing models is similar to the diagrammatic notation of the UML object diagrams; nothing has to be done by the modeler since this comes as part of the default environment of the platform. In order to customize the syntax and the editor behavior, a new \textit{View} must be created using the view manager at the upper-right corner of \circled{3} of Fig.~\ref{fig:gui}. A newly created view does not contain any \textit{ViewRule}; thus, the default style is still applied throughout the model elements in the editor. By defining a new \textit{ViewRule} \circled{4} for the selected elements in the graph \circled{5}, the designer can customize the way model elements are represented in the editor and how the user can interact with them accordingly. Figure~\ref{fig:4_template} shows how a sub-node (highlighted with a red dashed outline) is represented by means of a HTML/CSS fragment. It is worth noting that in such a fragment, the expression \texttt{\small \$\#\#name\$} specifies the system to search for the property \textit{name} in the modelling element associated with the current \textit{ViewRule}. However, this form of templating does not allow the specification of any positional feature. The user can only interact with the usual property dialog without performing direct manipulations of the modelling elements. 
In order to give some positional attribution to the modeling element, it has to be made \textit{measurable} as illustrated in Fig.~\ref{fig:4_measurable}. 
Once an element is made {"measurable"}, its movement and resizing capability may be further characterized by using the provided options; the element in figure has been constrained to be freely draggable, but only resizable through its lower-right corner.

\subsection{Positional constraint example}
\begin{figure}[h!]
    \centering
    \fbox{\includegraphics[width=0.95\linewidth]{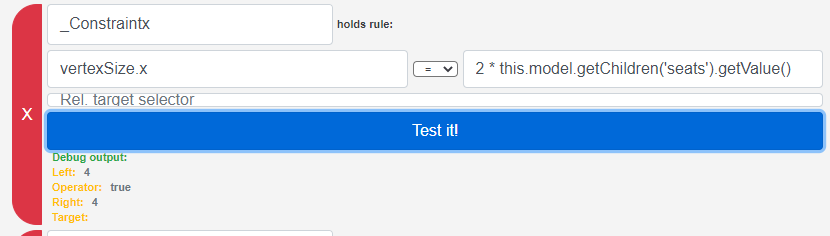}}
    \caption{A constraint action}
    \label{fig:4_m1_constraintx}
\end{figure}

Positional constraints are expressions that predicate the positioning of the modeling elements in the editor. In our airplane example, the number of \textit{seats} of an airplane is determined by its position along the x-axis. However, since the number of seats cannot be negative or greater than a reasonable upper limit, aircraft cannot be freely moved along the x-axis but only within a specific range. To this end, the following \textit{constraint} action (see Sect.~\ref{sec:positional_feature}) 
\begin{center}
    \texttt{\small vertexSize.x = 2 * this.model.getChildren('seats').getValue()}
\end{center}
can be provided in the dialog in Fig.~\ref{fig:4_m1_constraintx}.
Such a rule prescribes that X positioning of the aircraft must be double the amount of his passengers. If the actual position does not satisfy the predicate, then the underlying constraint solver will find the closest position satisfying the constraint, and the visual node will be moved accordingly. The constraint must be evaluated in correspondence to specific trigger events: when the airplane is created the first time, and when it is moved around. These two cases are captured by the \textit{onRefresh} and \textit{whileDragging} triggers that will be associated to the constraint above.

\begin{figure}
    \centering
    \fbox{\includegraphics[width=0.85\linewidth]{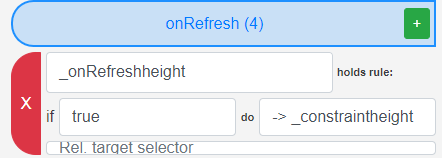}}
    \caption{A trigger condition}
    \label{fig:4_m1_trigger}
\end{figure}

%% file: related.tex
In this section, we discuss related work, which may be confined to the realm of {meta-editors}, {visual languages and positional grammars}, and general user interface and experience.
As to \textit{meta-editors}, we briefly surveyed existing tools and approaches in Sect.~\ref{sec:background}; the list of considered frameworks is not exhaustive, but most likely, they have pretty extended coverage of current meta-editor usage.
The corpus research related to \textit{visual language} is vast. In this context, Costagliola~\etal~\cite{Costagliola1998,874373} extend context-free grammars for visual languages by considering graphical constructs in addition to string concatenation. The paper represents a class of approaches that aim to support diagrammatic languages and generate the corresponding programming environment through so-called Visual Language Compiler-Compilers (VLCCs). In particular, relation-based representation describes a set of graphical objects and a set of relations. In contrast, a set of attributed graphical objects are specified by attribute-based representation. Here, the authors provide a first attempt to express language's attributes and references by graphical notation. Similar works, including \cite{10.1016/S1045-926X(05)80005-9,363633,275765}, include variants on the expressiveness and present formal characterizations. A more systematic review of the literature for the field of visual languages is deferred to future work. The main objective will be to identify approaches that provide insights and that can easily be leveraged to metamodeling. 
More recently, in~\cite{abrahao2017user} a discussion of how model-driven platforms fall short concerning \textit{User interface and eXperience} (UX). While most of the focus is on platform usability, characteristics, advantages, and disadvantages, very little is said on the intrinsic quality of the generated environments. Nevertheless, the paper introduces a notion of user experience for MDE (MX) to highlight the challenges and opportunities surrounding UX for MDE tool development, such as user model integration, processes for tailoring UX to a domain-specific language, and interoperability. In such a context, \tool makes a step forward in the usability of modeling tools by getting a more faithful alignment between domain community needs and provided modeling environments.

More general papers discussing what nowadays still is challenging in current platforms have been recently published (see~\cite{bucchiarone2021future,bucchiarone2020grand}). In particular, it is worth noting how accidental complexity is one of the major hindrances for broader and easier adoption of current model-driven tooling. The reasons are not easily identifiable but are related to the complexity of tools (e.g., Eclipse/EMF). Indeed, most of the time, the decision-making process concerning the platform design and implementation took place ages ago, with requirements and objectives mutated over the years.

%% file: 05-discussion.tex
A fundamental challenge for model-driven engineering is to reap the benefits of current generic platforms while addressing the need to design better domain-specific modeling environments. Despite the traction gained in the industry by MDE, such discipline is still affected by issues and problems as described in recent investigations~\cite{bucchiarone2021future,bucchiarone2020grand}.
For instance, a recurring complaint for industrial organizations considering the adoption of model-driven technologies is the lack of sufficient tool support~\cite{abrahao2017user}. In particular, current users are often exposed to a certain extent to forms of accidental complexity that became even more evident with the advent of low-code development platforms (LCDPs), where complexity is minimized, and usability~\cite{hassenzahl2006user} maximized. 

By designing and implementing \tool we were addressing the problem of making the modeler life a bit easier, i.e., we tried to borrow some of the ideas that made LCDPs successful, like zero setup, low or no maintenance, and advanced techniques for user interfaces. Notably, a prototype does not feature the technical maturity of a commercial or industrial grade tool but at least can help explore new territories and experiment with prospective solutions. In our case, the intention was to overcome the limitations due to the topological nature of the existing meta-editors (and also of the design choice of offering a limited set of pre-built functionalities). Many functionalities offered by \tool are obviously still requiring some polishing and simplified (and elegant) notations. While \tool features a strictly more expressive concrete syntax than the existing meta-editors, it is still unclear how far such expressiveness can go, e.g., it is unclear whether a sequence diagram editor can be easily arranged in \tool. Indeed, a formal characterization of the expressiveness of positional syntax is missing to the best of our knowledge. Nevertheless, the paper shows that enhancing what in \cite{abrahao2017user} is called \textit{user experience for MDE} (MX) by means of positional notations, the application landscape for MDE appears vaster and reasonably prone to new opportunities. As to \tool, it is a prototypical system that will be further developed and refined. Especially, the possibility of using it on the web represents a useful opportunity for those instructors who want to focus on the more conceptual and foundational aspects of modeling without spending ages in installing and configuring the n-th instance of the latest EMF bundle.



Besides improving the general maturity of the tool, future work includes a more foundational treatment of the class of positional syntax and editors; more specifically, how it differs from that of topological editors. Moreover, we would like to survey different engineering domains to assess whether model-driven engineering with the help of positional features, like the modeling tools for designing railway interlocking systems.

%% file: main.bbl
\begin{thebibliography}{10}
\providecommand{\url}[1]{#1}
\csname url@samestyle\endcsname
\providecommand{\newblock}{\relax}
\providecommand{\bibinfo}[2]{#2}
\providecommand{\BIBentrySTDinterwordspacing}{\spaceskip=0pt\relax}
\providecommand{\BIBentryALTinterwordstretchfactor}{4}
\providecommand{\BIBentryALTinterwordspacing}{\spaceskip=\fontdimen2\font plus
\BIBentryALTinterwordstretchfactor\fontdimen3\font minus
  \fontdimen4\font\relax}
\providecommand{\BIBforeignlanguage}[2]{{%
\expandafter\ifx\csname l@#1\endcsname\relax
\typeout{** WARNING: IEEEtran.bst: No hyphenation pattern has been}%
\typeout{** loaded for the language `#1'. Using the pattern for}%
\typeout{** the default language instead.}%
\else
\language=\csname l@#1\endcsname
\fi
#2}}
\providecommand{\BIBdecl}{\relax}
\BIBdecl

\bibitem{MDE}
D.~C. Schmidt, ``Model-driven engineering,'' \emph{Computer-IEEE Computer
  Society-}, vol.~39, no.~2, p.~25, 2006.

\bibitem{abrahao2017user}
S.~Abrah{\~a}o, F.~Bordeleau, B.~Cheng, S.~Kokaly, R.~Paige, H.~St{\"o}errle,
  and J.~Whittle, ``User experience for model-driven engineering: Challenges
  and future directions,'' in \emph{2017 ACM/IEEE 20th International Conference
  on Model Driven Engineering Languages and Systems (MODELS)}.\hskip 1em plus
  0.5em minus 0.4em\relax IEEE, 2017, pp. 229--236.

\bibitem{GMF}
\BIBentryALTinterwordspacing
``{Eclipse project: Graphical Modeling Framework (GMF)}.'' [Online]. Available:
  \url{https://www.eclipse.org/modeling/gmp/}
\BIBentrySTDinterwordspacing

\bibitem{kolovos2017eugenia}
D.~S. Kolovos, A.~Garc{\'\i}a-Dom{\'\i}nguez, L.~M. Rose, and R.~F. Paige,
  ``Eugenia: towards disciplined and automated development of gmf-based
  graphical model editors,'' \emph{Software \& Systems Modeling}, vol.~16,
  no.~1, pp. 229--255, 2017.

\bibitem{SIRIUS}
\BIBentryALTinterwordspacing
``{Eclipse project: Sirius}.'' [Online]. Available:
  \url{https://www.eclipse.org/sirius/}
\BIBentrySTDinterwordspacing

\bibitem{takahashi1991general}
S.~Takahashi, S.~Matsuoka, A.~Yonezawa, and T.~Kamada, ``A general framework
  for bi-directional translation between abstract and pictorial data,'' in
  \emph{Proceedings of the 4th annual ACM symposium on User interface software
  and technology}, 1991, pp. 165--174.

\bibitem{smolander1991metaedit}
K.~Smolander, K.~Lyytinen, V.-P. Tahvanainen, and P.~Marttiin, ``Metaedit—a
  flexible graphical environment for methodology modelling,'' in
  \emph{International Conference on Advanced Information Systems
  Engineering}.\hskip 1em plus 0.5em minus 0.4em\relax Springer, 1991, pp.
  168--193.

\bibitem{EMF}
F.~Budinsky, D.~Steinberg, E.~Merks, R.~Ellersick, and T.~Grose, \emph{{Eclipse
  Modeling Framework}}.\hskip 1em plus 0.5em minus 0.4em\relax Addison Wesley,
  2003.

\bibitem{basciani2014mdeforge}
F.~Basciani, J.~Di~Rocco, D.~Di~Ruscio, A.~Di~Salle, L.~Iovino, and
  A.~Pierantonio, ``Mdeforge: an extensible web-based modeling platform,'' in
  \emph{2nd International Workshop on Model-Driven Engineering on and for the
  Cloud, CloudMDE 2014, Co-located with the 17th International Conference on
  Model Driven Engineering Languages and Systems, MoDELS 2014}, vol.
  1242.\hskip 1em plus 0.5em minus 0.4em\relax CEUR-WS, 2014, pp. 66--75.

\bibitem{france2006repository}
R.~France, J.~Bieman, and B.~H. Cheng, ``Repository for model driven
  development (remodd),'' in \emph{International Conference on Model Driven
  Engineering Languages and Systems}.\hskip 1em plus 0.5em minus 0.4em\relax
  Springer, 2006, pp. 311--317.

\bibitem{Costagliola1998}
\BIBentryALTinterwordspacing
G.~Costagliola, A.~De~Lucia, S.~Orefice, and G.~Tortora, \emph{Positional
  Grammars: A Formalism for LR-Like Parsing of Visual Languages}.\hskip 1em
  plus 0.5em minus 0.4em\relax New York, NY: Springer New York, 1998, pp.
  171--191. [Online]. Available:
  \url{https://doi.org/10.1007/978-1-4612-1676-6_5}
\BIBentrySTDinterwordspacing

\bibitem{874373}
G.~Costagliola and G.~Polese, ``Extended positional grammars,'' in
  \emph{Proceeding 2000 IEEE International Symposium on Visual Languages},
  2000, pp. 103--110.

\bibitem{10.1016/S1045-926X(05)80005-9}
\BIBentryALTinterwordspacing
E.~J. Golin, ``Parsing visual languages with picture layout grammars,''
  \emph{J. Vis. Lang. Comput.}, vol.~2, no.~4, p. 371–393, Dec. 1991.
  [Online]. Available: \url{https://doi.org/10.1016/S1045-926X(05)80005-9}
\BIBentrySTDinterwordspacing

\bibitem{363633}
K.~Marriott, ``Constraint multiset grammars,'' in \emph{Proceedings of 1994
  IEEE Symposium on Visual Languages}, 1994, pp. 118--125.

\bibitem{275765}
K.~Wittenburg, ``Earley-style parsing for relational grammars,'' in
  \emph{Proceedings IEEE Workshop on Visual Languages}, 1992, pp. 192--199.

\bibitem{bucchiarone2021future}
A.~Bucchiarone, F.~Ciccozzi, L.~Lambers, A.~Pierantonio, M.~Tichy, M.~Tisi,
  A.~Wortmann, and V.~Zaytsev, ``What is the future of modeling?'' \emph{IEEE
  software}, vol.~38, no.~2, pp. 119--127, 2021.

\bibitem{bucchiarone2020grand}
A.~Bucchiarone, J.~Cabot, R.~F. Paige, and A.~Pierantonio, ``Grand challenges
  in model-driven engineering: an analysis of the state of the research,''
  \emph{Software and Systems Modeling}, vol.~19, no.~1, pp. 5--13, 2020.

\bibitem{hassenzahl2006user}
M.~Hassenzahl and N.~Tractinsky, ``User experience-a research agenda,''
  \emph{Behaviour \& information technology}, vol.~25, no.~2, pp. 91--97, 2006.

\end{thebibliography}
